\documentclass[final,5p,times,twocolumn]{elsarticle}

\usepackage{graphicx}
\usepackage{lineno}
\usepackage{amssymb}
\journal{Nuclear Instruments and Methods A}

\begin{document}
\linenumbers

\begin{frontmatter}

%% Title, authors and addresses

%% use the tnoteref command within \title for footnotes;
%% use the tnotetext command for theassociated footnote;
%% use the fnref command within \author or \address for footnotes;
%% use the fntext command for theassociated footnote;
%% use the corref command within \author for corresponding author footnotes;
%% use the cortext command for theassociated footnote;
%% use the ead command for the email address,
%% and the form \ead[url] for the home page:
%% \title{Title\tnoteref{label1}}
%% \tnotetext[label1]{}
%% \author{Name\corref{cor1}\fnref{label2}}
%% \ead{email address}
%% \ead[url]{home page}
%% \fntext[label2]{}
%% \cortext[cor1]{}
%% \address{Address\fnref{label3}}
%% \fntext[label3]{}

\title{Achievements of the ATLAS Upgrade Planar Pixel Sensors R\&D Project }

% if there is only one institution, use this form:
\author{G. Calderini \\ (on behalf of the ATLAS Planar Pixel Sensors Collaboration)}
\address{Laboratoire de Physique Nucl\'eaire et des Hautes Energies (LPNHE) Paris, France \\ Dipartimento di Fisica E. Fermi, Universit\'a di Pisa, Pisa, Italy}

\begin{abstract}
This paper reports on recent accomplishments and ongoing work of the ATLAS Planar Pixel Sensors R\&D project. Special attention is given in particular to new testbeam results obtained with highly irradiated sensors, developments in the field of slim and active edges and first steps towards prototypes of future pixel modules.

\end{abstract}

\begin{keyword}
Planar Silicon radiation detectors, tracking detectors, fabrication technology

%% PACS codes here, in the form: \PACS code \sep code
%% Find PACS codes here: http://www.aip.org/pacs/pacs2010/individuals/pacs2010_regular_edition/index.html

%% MSC codes here, in the form: \MSC code \sep code
%% or \MSC[2008] code \sep code (2000 is the default)

\end{keyword}

\end{frontmatter}

%% \linenumbers

%% main text
\section{Introduction}

To extend the physics reach of the LHC, accelerator upgrades are planned which will increase the integrated luminosity to beyond 3000 fb$^{-1}$ and the pile-up per bunch-crossing by a factor 5 to 10. To cope with the increased occupancy and radiation damage, the ATLAS experiment plans to introduce an all-silicon inner tracker with the HL-LHC upgrade. To investigate the suitability of pixel sensors using the proven planar technology for the upgraded tracker, the ATLAS Upgrade Planar Pixel Sensor R\&D Project (PPS) was established comprising 19 institutes and more than 80 scientists. Main areas of research are the performance assessment of planar pixel sensors with different designs and substrate thicknesses up to the HL-LHC fluence, the achievement of slim or active edges to provide low geometric inefficiencies without the need for shingling of modules and the exploration of possibilities for cost reduction to enable the instrumentation of large areas. This paper gives an overview of recent accomplishments and ongoing work of the R\&D project. 

\section{The planar pixels sensor technology}
Planar pixel sensors are widely used since many years in the tracking detectors of many high-energy experiments. This is now a mature and sound technology, allowing to reach a very high production yield at a limited cost. The experience achieved in the construction and in the many years of operations of detectors equipped with planar pixel sensors represent a guarantee for future trackers. At the same time, a few axes of development can be followed to achieve further improvements. Radiation hardness has been increased, currently allowing good operation conditions even after a fluence of a few $10^{15} n_{eq}/cm^2$ which is what is expected for the non-innermost pixel layers during the Phase-II of the LHC upgrade. Special processes are under study to allow an increased geometrical efficiency by reducing the size of the dead region at the edge of the sensors. This is critical in the assembly of the staves, where the modules are tiled, or for layers in which there is not enough room to stagger the staves in the $r-\phi$ plane to allow an overlap of the corresponding active regions. A few options, as larger wafers, multi-sensor modules or processes as the n-in-p, are also being developed to further reduce the cost of the productions.       

\section{Development of active-edge sensors}
\begin{figure}[h!] 
\centering 
\includegraphics[width=0.9\columnwidth,keepaspectratio]{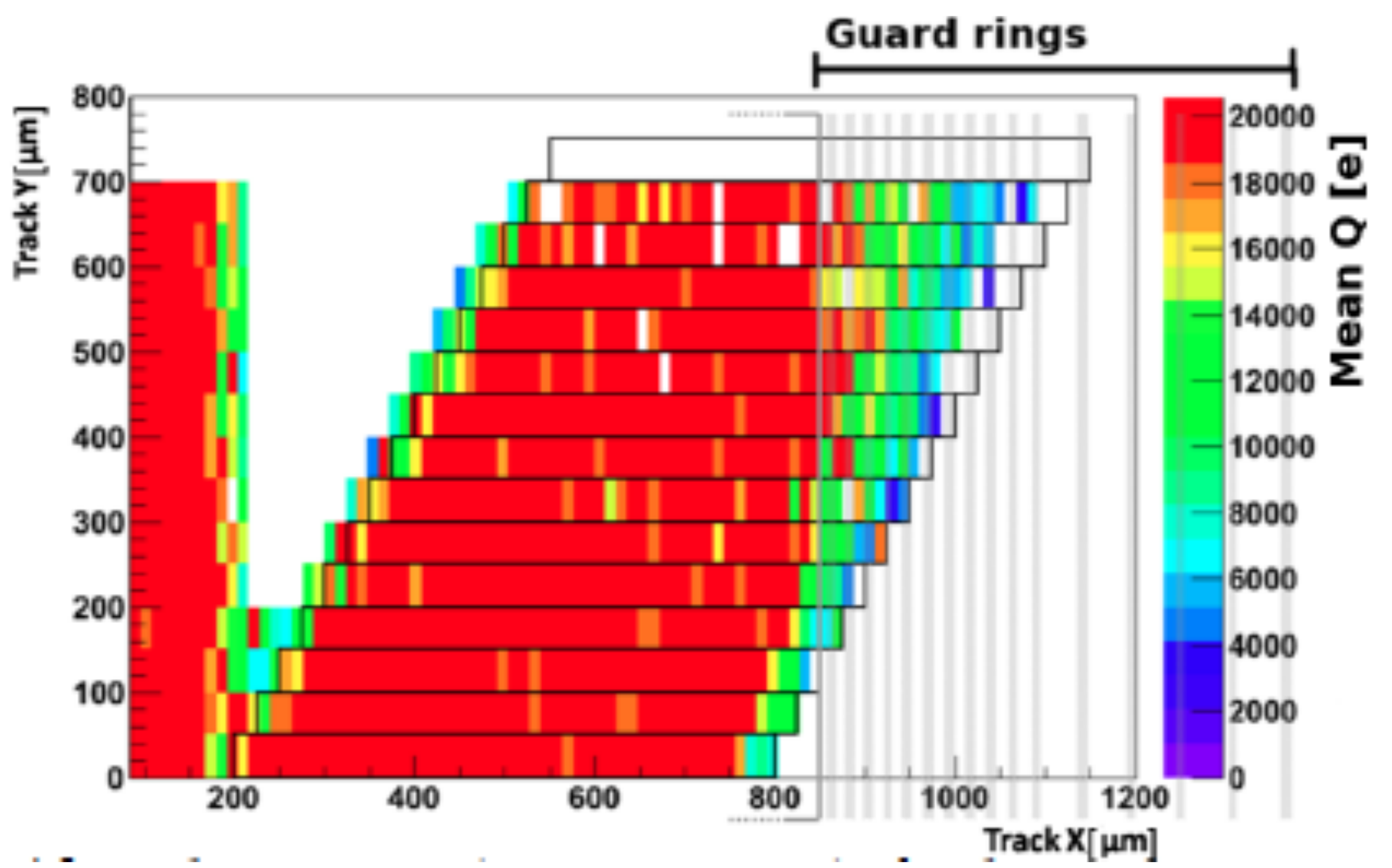}
\caption{Beam-test analysis of special sensors developed to study the layout optimisation for the ATLAS IBL. Each horizontal rectangle represents a block of pixels and each block is pushed at different positions below the guard rings, which are implanted on the opposite face.}
\label{fig:IBL}
\end{figure}

The need to reduce as much as possible the size of the dead region at the border of the sensors has driven the planar pixel community to use intensely the device simulation tools available and optimise the sensor layout. Good results have been achieved by reducing the guard ring region, which represents a low-efficiency portion of the sensor due to the lower electrical field and the distance from the first row of pixels. Already for the design of the ATLAS Insertable B-Layer sensors, preliminary simulations (see for instance \cite{hiroshima_2009}) had indicated that the number of guard rings could be reduced and the n-in-n nature of the sensors, in which pixels and guard rings are on opposite faces, could allow to push the first row of pixels inside the guard-ring area. Beam-test analysis showed that some charge collection efficiency was still possible for particles crossing the detector in this region (Fig. \ref{fig:IBL}), thus allowing an improvement of the geometrical acceptance. The solution has been indeed adopted for the ATLAS IBL sensors.  

\begin{figure}[h!] 
\centering 
\includegraphics[width=0.9\columnwidth,keepaspectratio]{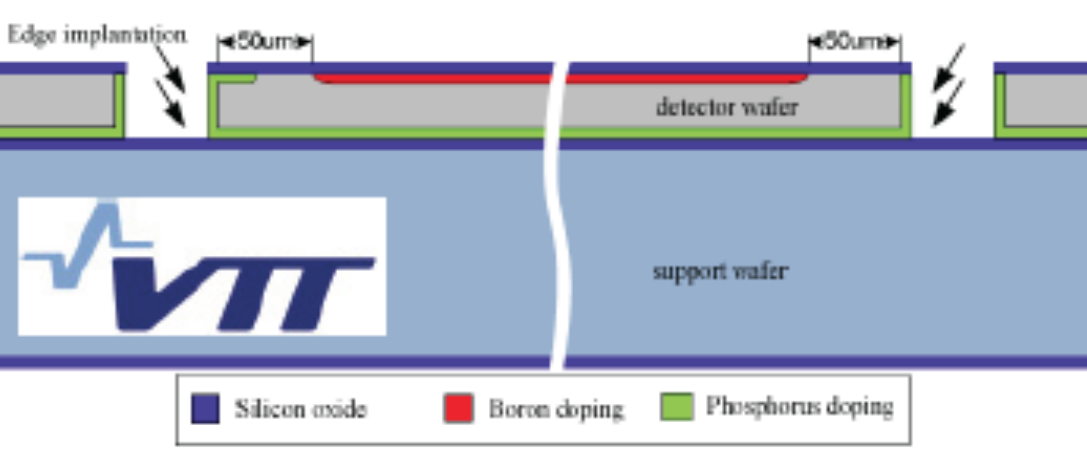}
\caption{Sketch of the VTT process to achieve active-edge sensors}
\label{fig:vtt-active-edge}
\end{figure}

More recently, further optimisation in the number of the guard-rings has been achieved in both n-in-n and n-in-p productions, allowing to reduce the size of the inefficient region down to a typical scale of $300-400\mu m$ (see for example \cite{liverpool-gr-reduction}). In addition, an alternative approach based on the use of DRIE (Deep Reactive Ion Etching) and SCP (Scribe, Cleave, Passivate) techniques has allowed the construction of active-edge devices. In the first case the method consists in producing sensors with an edge doping of the same type with respect to the back-side. The net result is to have the cut-line inside an equipotential region, where the absence of electrical field prevents the generation of edge surface current. Different processes allow this result. A production of n-in-p devices at VTT \cite{vtt, mpp-active-edge} uses DRIE to excavate a deep and large trench in order to expose the side. A lateral implantation is then started so that the doping of the vertical region becomes similar to that of the backside (see Fig. \ref{fig:vtt-active-edge}).     

\begin{figure}[h!] 
\centering 
\includegraphics[width=0.9\columnwidth,keepaspectratio]{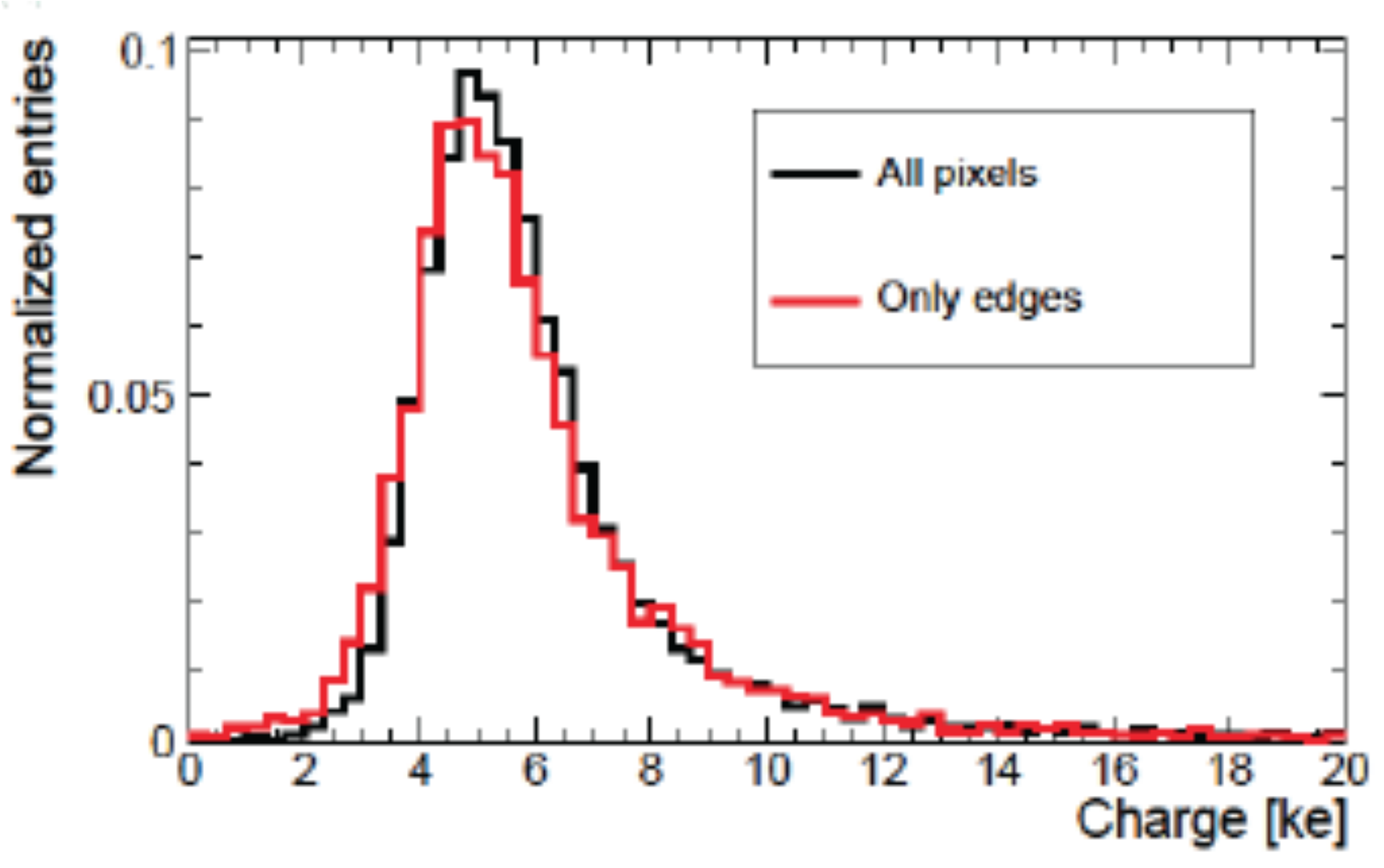}
\caption{Beam test analysis of collected charge for VTT active-edge sensors. The inclusive distribution and the one for the pixels of the first row are very similar.}
\label{fig:vtt-cce}
\end{figure}

Using this technique, $100~\mu m$ and $200~\mu m$-thick n-in-p sensors are produced in collaboration with MPP Munich, with a distance of the first pixel from the edge of the order of $50$ and $125 \mu m$. Beam test results indicate that the collected charge distribution is identical for the pixels of the first row and the others, thus showing that the concept works (see Fig. \ref{fig:vtt-cce}).

The beam-test analysis indicates that a hit efficiency of $84 ^{+9}_{-14}\%$ is achieved in the last $50~\mu m$ of the sensor edge, beyond the last pixel implant.

\begin{figure}[h!] 
\centering 
\includegraphics[width=0.9\columnwidth,keepaspectratio]{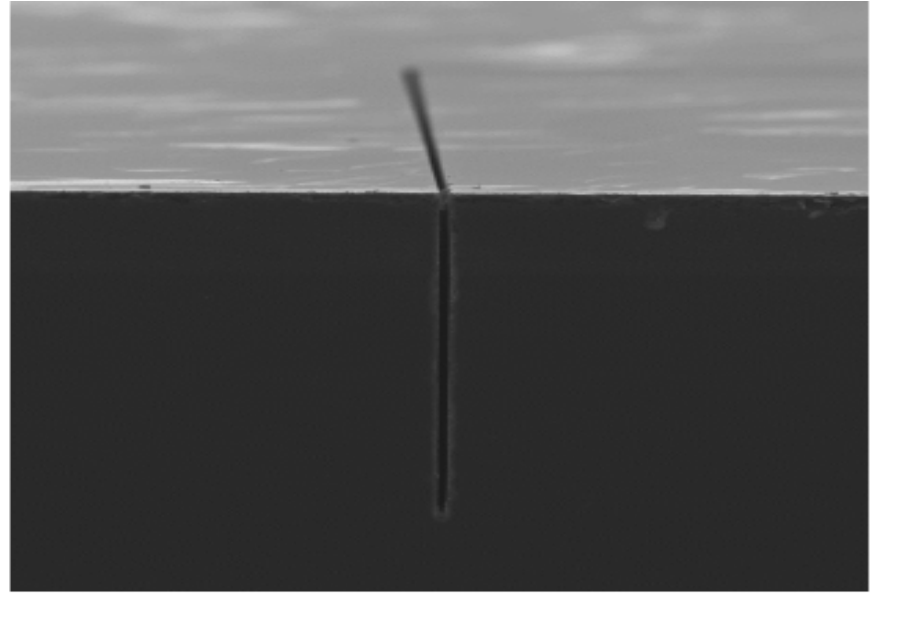}
\caption{Detail of a deep trench produced by DRIE on a test bulk at FBK}
\label{fig:fbk-active-edge}
\end{figure}

A similar method is used in the FBK (Fondazione Bruno Kessler) \cite{fbk} process. In this case, the deep trench is again excavated by DRIE, but the lateral doping is obtained by diffusion instead of implantation. The doped trench is finally filled with poly-silicon \cite{fbk-active-edge, cg-active-edge}. The uniformity of the poly-silicon filling is a critical phase of the process. Residuals of air in the $200~\mu m$-deep and $4.5~\mu m$-wide trench (Fig. \ref{fig:fbk-active-edge}) could severely damage the device during the high-temperature phases of the process. 

\begin{figure}[h!] 
\centering 
\includegraphics[width=0.58\columnwidth,keepaspectratio]{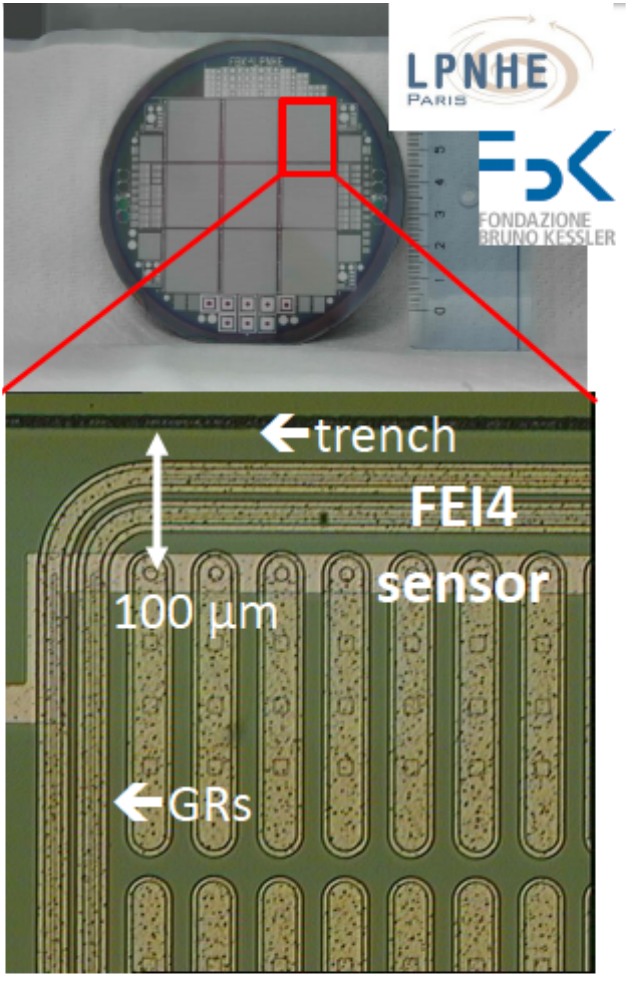}
\caption{Detail of the edge region of a 2-guard-ring FE-I4 geometry pixel sensor.}
\label{fig:fbk-picture}
\end{figure}

Using this technique, $200~\mu m$ thick n-in-p sensors are produced by FBK in collaboration with LPNHE Paris. ATLAS FE-I3 and FE-I4 designs \cite{fe-i4} are used, with different guard-ring numbers and edge distance configurations. The distance of the first row of pixels from the cut-line is typically of the order of $100-200~\mu m$ (Fig. \ref{fig:fbk-picture}). 

\begin{figure}[h!] 
\centering 
\includegraphics[width=0.75\columnwidth,keepaspectratio]{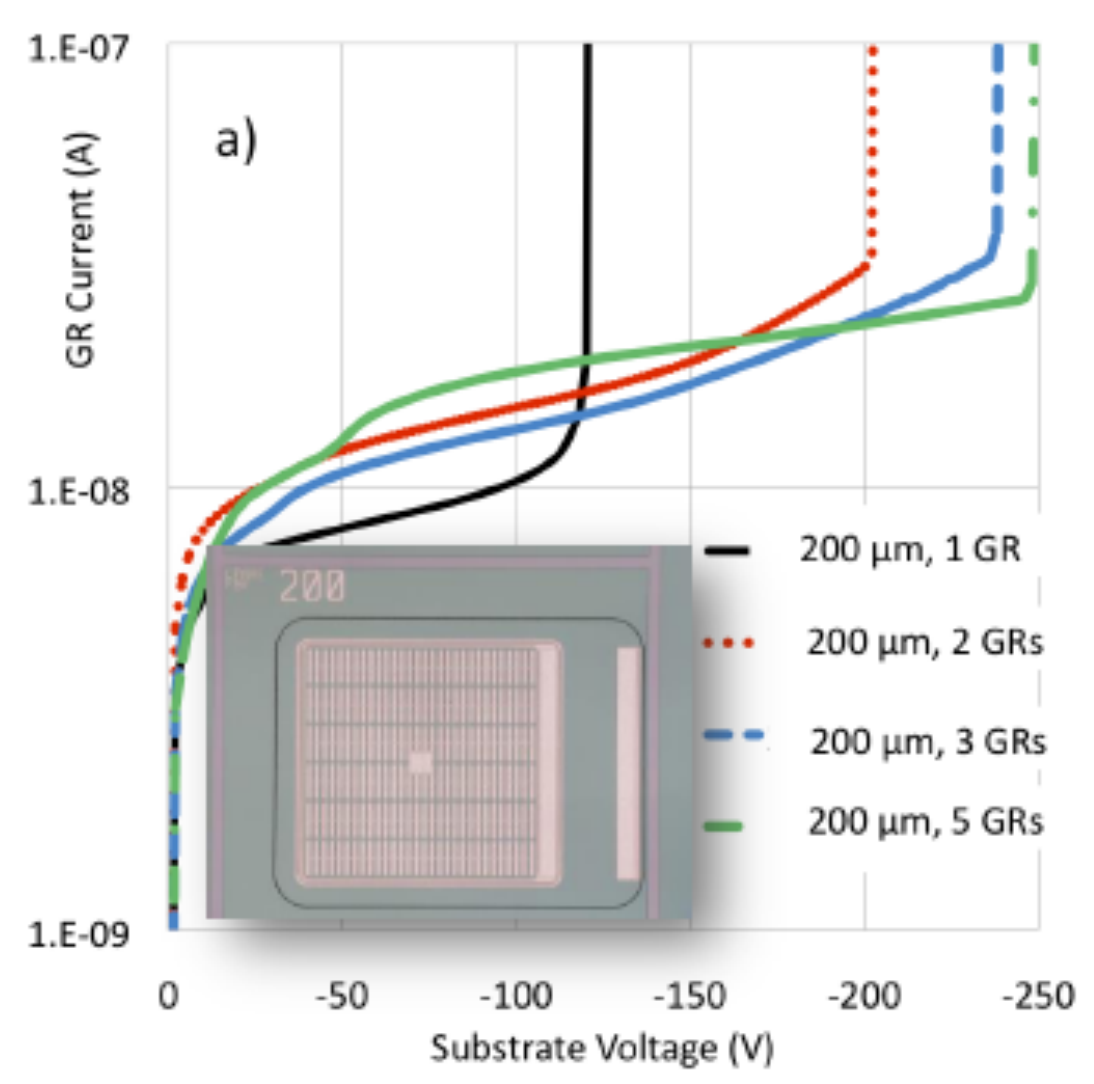}
\caption{Typical IV behaviour of IV curves for baby detectors with 1, 2, 3, and 5 guard rings. The curves are in very good qualitative agreement with simulations.}
\label{fig:fbk-IV}
\end{figure}

Baby-detectors with a reduced number of pixels but the same guard-ring and edge configuration as the main structures are also produced on the same wafers for test purpose. Measurements show a breakdown voltage exceeding the 100V (even 200V for sensors with more than 1 guard ring), in excellent agreement with simulations (see Fig. \ref{fig:fbk-IV}). 

\begin{figure}[h!] 
\centering 
\includegraphics[width=0.75\columnwidth,keepaspectratio]{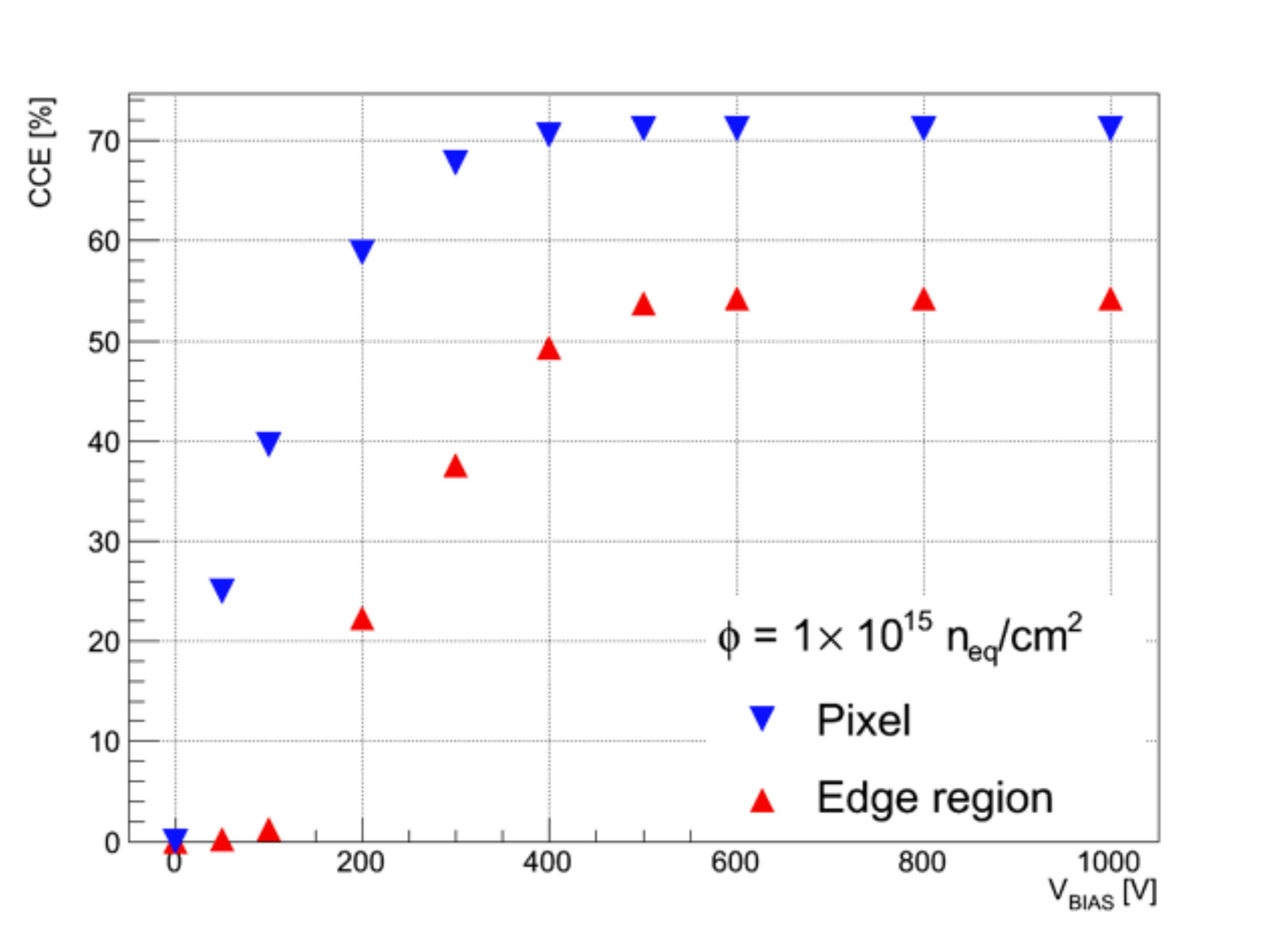}
\caption{Expected charge collection efficiency after a $10^{15} n_{eq}/cm^{2}$ dose as a function of bias voltage for central and edge pixels (simulation) .}
\label{fig:fbk-cce}
\end{figure}

Simulations indicate that even at a dose of $10^{15} n_{eq}/cm^2$ the charge collection efficiency of the pixels of the first row is still significant with respect to the pixels of the central region, provided that a bias voltage exceeding 2-300V is applied (see Fig. \ref{fig:fbk-cce}).  

The sensors are presently being bump-bonded to ATLAS FE-I4 chips to be analysed at beam-tests. 

\begin{figure}[h!] 
\centering 
\includegraphics[width=0.75\columnwidth,keepaspectratio]{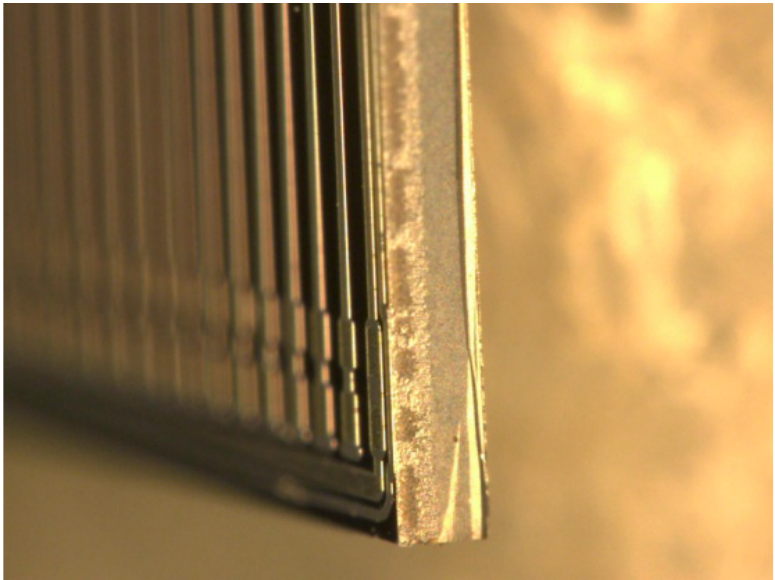}
\caption{Photo of the sensor sidewall after cleaving and passivation phase.}
\label{fig:scipp-scp}
\end{figure}

A different approach to the problem of edge current has been tested by SCIPP (UCSC) \cite{scipp} in collaboration with U.S. Naval Research Lab (NRL) \cite{nrl}. A Scribe-Cleave-Passivate (SCP) technique has been used to block the edge current in the sensors \cite{scp, scp2}. In this approach the detector is scribed along the edge by laser or XeF$_2$ etching and cleaved. Once the edge is exposed, the sensor undergoes a passivation phase via plasma-enhanced CVD or alumina deposition 
(see Fig. \ref{fig:scipp-scp}). 

\begin{figure}[h!] 
\centering 
\includegraphics[width=0.9\columnwidth,keepaspectratio]{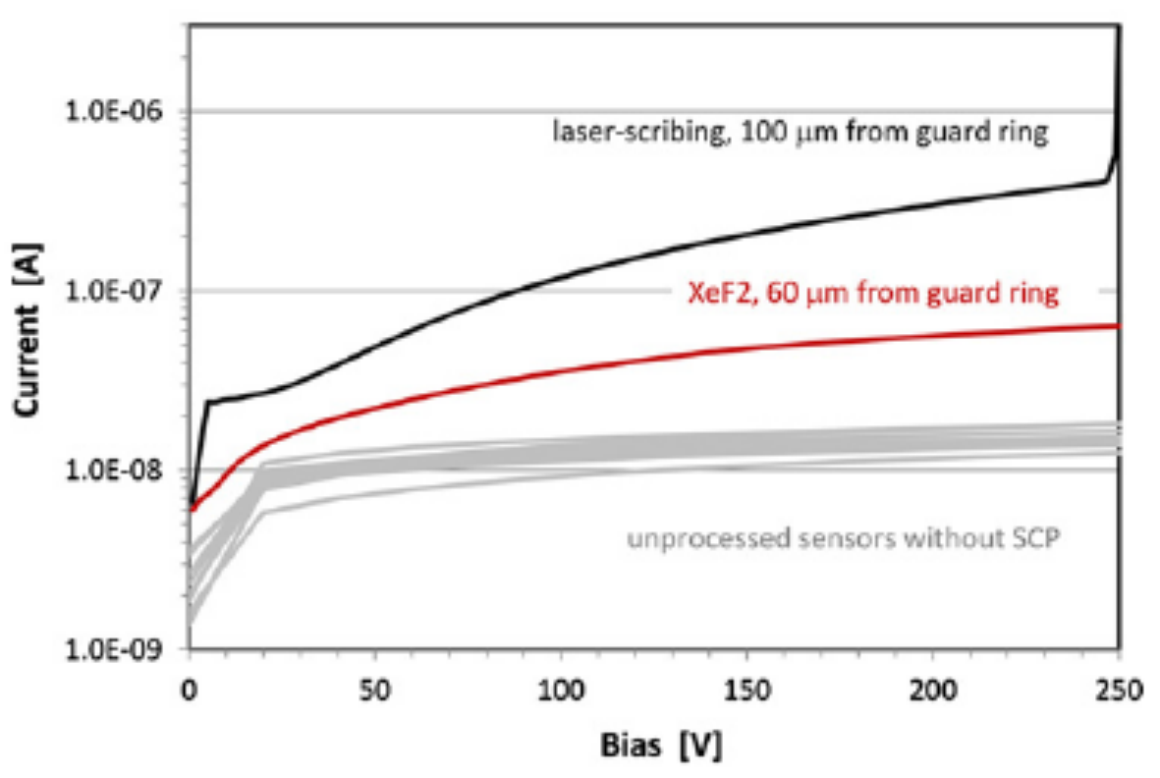}
\caption{Leakage current behaviour for sensors scribed with a laser (more damaging) at a distance of $100 \mu m$ from the guard ring and with XeF$_2$ PECVD (less damaging) at a distance of $60 \mu m$. In the second case, in spite of the reduced distance, the current is lower due to the better quality of the scribing}
\label{fig:scipp-IV}
\end{figure}

The fixed interface charge determined by the passivation process allows to control the potential across the sidewall, minimising the edge current (see Fig. \ref{fig:scipp-IV}) \cite{vitaliy}.      

Further investigations are under way to determine the behaviour of the sensors treated with SCP after radiation damage.

\section{Reducing the costs}
One of the key activities in the recent development of planar pixel sensors has been the attempt to reduce the cost of productions. This is critical whenever the use of planar pixels is targeted for middle- and large-surface detectors, as is the case for the external layers of a tracker. Several axes of research have been developed.

\subsection{Cheaper interconnections} 
Significant R\&D has been devoted to reduce the price of sensor-FEE interconnections per unit area, by replacing the standard bump-bonding by cheaper technologies.
In the framework of the ATLAS PPS project, MPP Munich in collaboration with Fraunhofer EMFT \cite{fraunhofer} has investigated the possible use of Solid-Liquid Inter-Diffusion (SLID) process \cite{mpp-active-edge}. The smaller number of process steps with respect to bump-bonding could translate in reduced cost. In addition, the limitations on the geometrical pitch are less severe in this technology. Wafer to wafer as well as chip to wafer connections are feasible. SLID-based modules are expected to be delivered by the end of 2013 and they will be characterised with beam-tests.

Alternative low-cost bonding technologies as the IBM C4NP \cite{c4np} are also being evaluated by the ATLAS PPS Collaboration.       

\subsection{Multi-chip modules} 
\begin{figure}[h!] 
\centering 
\includegraphics[width=0.6\columnwidth,keepaspectratio]{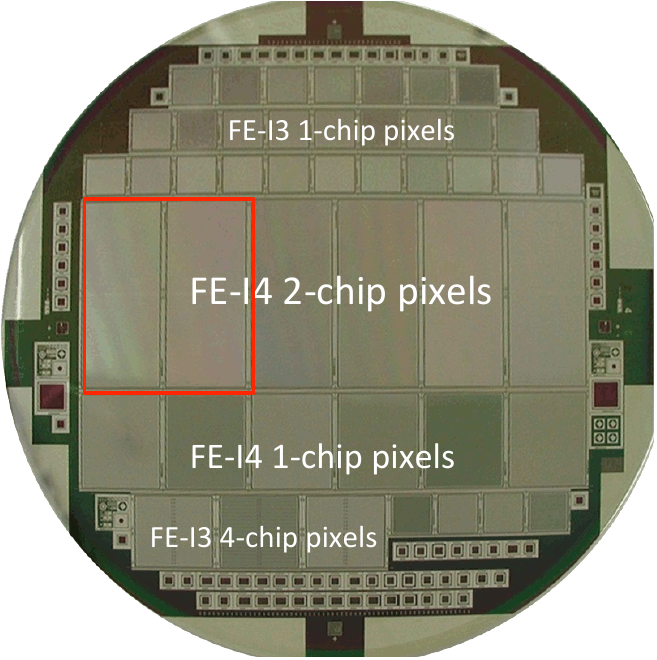}
\caption{Layout of a recent 6-inch production by Hamamatsu Photonics KK, in collaboration with the ATLAS Japan Silicon group. The large 2x2 FE-I4-design sensors are visible in the central part of the wafer, together with FE-I4-design and FE-I3-design single sensors.}
\label{fig:hpk}
\end{figure}

\begin{figure}[h!] 
\centering 
\includegraphics[width=0.8\columnwidth,keepaspectratio]{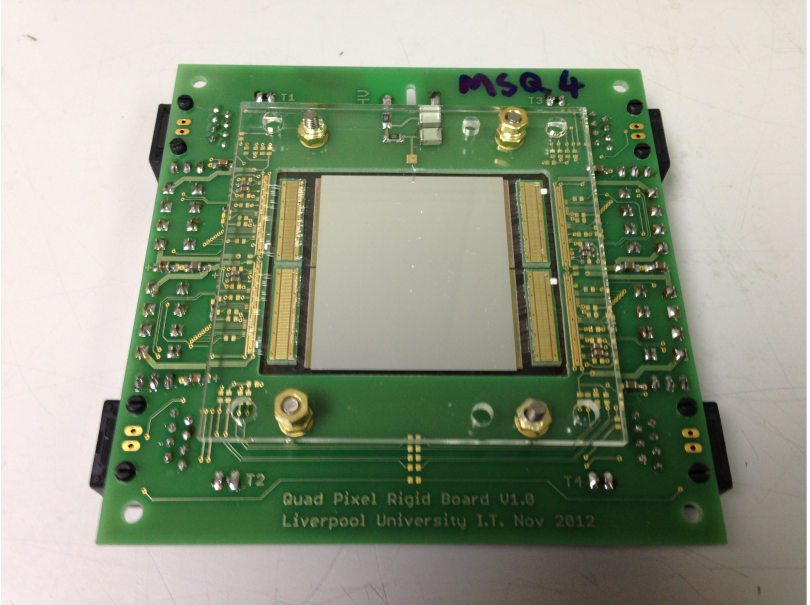}
\caption{Detail of a large 2x2 sensor produced by Micron in collaboration with the ATLAS Liverpool and Glasgow groups. The sensors is already bump-bonded and mounted on the periphery card to be read-out and installed in the beam-test.}
\label{fig:micron}
\end{figure}

An effective solution to achieve lower-cost productions is the development of larger sensors, which is made possible by the use of larger wafers by foundries, and the assembly of multi-chip modules. Many sensor providers recently moved their productions to 6-inch and even 8-inch wafers. This allows the optimisation of the wafer layout and the presence of more large-area sensors on the same wafer, features which translate globally into a significant cost-reduction. On the other hand, the larger size of the sensors allows the assembly of multi-chip modules, with a consequent reduction of the number of handling operations per unit of area and a significant cost reduction. These larger sensors can fit well the specifications for outer layers and the side end-cap regions.
In the framework of the ATLAS PPS Collaboration, recent productions of large "quad" sensors on 6-inch wafers have been done at CiS \cite{cis}, HPK (see Fig. \ref{fig:hpk}) \cite{hpk} and Micron (see Fig. \ref{fig:micron}) \cite{micron}. 
Some of the multi-chip modules have already been measured at beam-tests and the analysis of the results is underway.

\section{Design improvements}

\begin{figure}[h!] 
\centering 
\includegraphics[width=0.7\columnwidth,keepaspectratio]{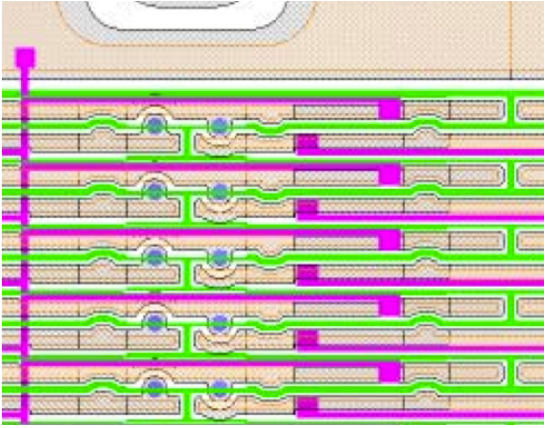}
\caption{Detail of the poly-silicon biasing scheme for a sensor with $25 x 500 \mu m$ pixels designed by the ATLAS Japan Silicon group and produced by Hamamatsu Photonics KK. The biasing lines are staggered on neighboring pixels.}
\label{fig:kek-bias}
\end{figure}

In addition to the cost reduction, an important line of research consists in the further optimisation of the sensor design to achieve the best possible performance even after heavy irradiation and years of operation in a high-energy physics experiment. One of the critical issues with the punch-through structure of the biasing system is a certain loss of efficiency in the dot region after irradiation. An effort has been made to provide different biasing schemes in order to avoid such a problem. The design of bias paths with poly-silicon structures has been studied by the ATLAS Japan Silicon group, which has proposed a number of solutions over the years. The position of the poly-silicon line inside the pixel is of great importance since it can affect the charge collection efficiency. A new design has been recently proposed in which the line runs along the inner part of the pixel edge, which seems to optimise the hit efficiency, even after radiation damage. In addition a new proposal of biasing scheme for $25 x 500 \mu m$ pixels has been proposed. In this design the bias lines are staggered on alternating pixels (see Fig. \ref{fig:kek-bias}).

\begin{figure}[h!] 
\centering 
\includegraphics[width=0.7\columnwidth,keepaspectratio]{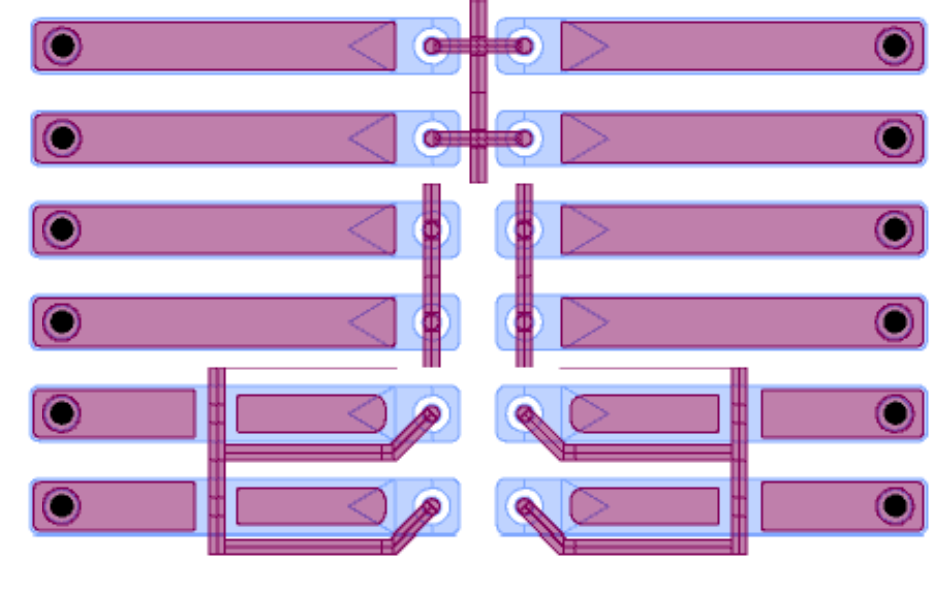}
\caption{Detail of the metal biasing scheme for a sensor with $50 x 250 \mu m$ pixels designed by the ATLAS Dortmund group and produced by CiS. Different geometry designs have been tested to optimise the sensor performance.}
\label{fig:dortmund-bias}
\end{figure}

A different technique has been proposed by the ATLAS group of Dortmund, which studied a distribution scheme obtained by bias rails, narrow metal strips running over the oxide layer and providing the electrical network to the pixels. Also in this case the geometrical configuration can influence the hit efficiency and several different concepts have been tested (see Fig. \ref{fig:dortmund-bias}). 
 
\section{Bulk materials}
Special mention needs to be done for a recent study by the ATLAS Dortmund group, which is trying to improve the radiation hardness of n-in-n sensors by the use of a MCz bulk. The intermediate layers of the upgraded ATLAS tracker will be exposed to a mix of ionising and non-ionising radiation. MCz material is expected to have a better performance with respect to standard DOFZ bulk under these conditions \cite{kramb}. For this reason, MCz FE-I4-design prototype sensors have been produced and irradiated with protons at the CERN PS up to three different fluences (0.8, 1.6 and 3.0 $10^{15} n_{eq}/cm^2$. After this first phase they have been tested with a $^{90}$Sr source and at a beam-test at CERN, showing a good residual charge collection efficiency, at least for the first two irradiation samples.  

They are presently undergoing the second phase of irradiation, the one involving neutrons, at the Ljubljana reactor. The plan is to test them again in one of the next PPS beam-test. 
    
\section{Conclusions} 
Planar pixel sensors represent a sound and well-established technology for the tracking detectors of high-energy physics experiments. 
Many activities are developed in parallel in the framework of the ATLAS PPS Collaboration to improve the sensors performance, even after the heavy irradiations expected in the Phase-II of LHC, and to reduce the costs of productions. New developments in techniques of fabrication of active-edge sensors have been presented. 
Progress in interconnection technology allows to reduce the cost of module productions, as well as the use of larger sensors and multi-chip modules. New solutions for the biasing schemes have been presented.

\section{Acknowledgements}
This work has been partially performed in the framework of the CERN RD50 Collaboration. Some of the irradiations and the beam test measurements
leading to these results has received funding from the European Commission under the FP7 Research Infrastructures project AIDA, grant agreement no. 262025.
The edgeless sensor production at FBK was supported in part by the Autonomous Province of Trento, Project MEMS2.

\end{document}